\documentclass[10pt, a4paper, journal, two side]{ICTR}
\usepackage[utf8,nocaptions]{vietnam}

\usepackage{graphicx}
\usepackage[table]{xcolor}
\usepackage{epstopdf} 
\usepackage{subfigure}
\usepackage[ruled,linesnumbered]{algorithm2e}

\usepackage{multirow}
\usepackage{tabularx}
\usepackage{makecell}
\usepackage{booktabs}

\usepackage{textcomp}
\usepackage{newtxmath}

\usepackage{cite}
\usepackage[square, comma, numbers, sort&compress]{natbib}

\usepackage{listings} 
\usepackage{lineno}
\usepackage{balance}

\begin{document}
\title{Impact of Inter-Channel Interference on Shallow Underwater Acoustic OFDM Systems}
\AuthorName     	{Do Viet Ha$^1$, Nguyen Tien Hoa$^2$, Nguyen Van Duc$^{2}$}
\AuthorAddress  	{$^1$ Faculty of Electrical and Electronic Engineering, University of Transport and Communications, Hanoi, Vietnam\\
 			     		$^2$ School of Electronics and Telecommunications, Hanoi University of Science and Technology, Hanoi, Vietnam}

\Correspondence	{Nguyen Van Duc, duc.nguyenvan1@hust.edu.vn}


\maketitle

\setcounter{page}{1}

\begin{abstract}
	This paper investigates the impacts of Inter-Channel Interference (ICI) effects on a shallow underwater acoustic (UWA) orthogonal frequency-division multiplexing (OFDM) communication system. Considering both the turbulence of the water surface and the roughness of the bottom, a stochastic geometry-based channel model utilized for a wide-band transmission scenario has been exploited to derive a simulation model. Since the system bandwidth and the sub-carrier spacing is very limited in the range of a few kHz, the channel capacity of a UWA system is severely suffered by the ICI effect. For further investigation, we construct the signal-to-noise-plus-interference ratio (SINR) based on the simulation model, then evaluate the channel capacity. Numerical results show that the various factors of a UWA-OFDM  system as subcarriers, bandwidth, and OFDM symbols \textcolor{black}{affect} the channel capacity under the different Doppler frequencies. Those observations give hints to select the good parameters for UWA-OFDM systems.
\end{abstract}

\begin{keywords}
	Underwater acoustic (UWA) , Geometry-based Channel modeling, Underwater OFDM systems
\end{keywords}

\section{\bfseries Introduction}\label{sec:intro}
	
Underwater information systems have many applications in commercial and military \cite{Marani2009}. Nowadays, the development and improvement of the service quality of the system is facing many challenges such as frequency spectrum limitation, time and frequency-dependent channel characteristics \cite{Kim2015, Hoa2013}. There is a fact that the velocity of water sound waves is much lower than the velocity of electromagnetic waves, and therefore the prominent features of the UWA channels are a large \textcolor{black}{delay spread} and strong Doppler effects \cite{Ebihara2012}. The OFDM technique is widely used thanks to its ability in eliminating the inter-symbol interference (ISI) which is caused by a large delay spread \cite{Tao2018, Zhang2013}. However, OFDM systems are very easily influenced by the Doppler shifts that result in the so-called inter-carrier interference (ICI) and degrade the system performance dramatically \cite{MinhHai2015}.

In shallow water environments, acoustic communication systems \textcolor{black}{are suffered from} the Doppler shifts with the two main sources, which are the relative movement between the transmitter/the receiver and \textcolor{black}{the} disturbance from the water surface \cite{Yoshizawa2016,ha2016proposals}. In particular, a surface displacement process usually varies very fast over time, thus an unpredictable Doppler frequency shift creates many difficulties in identifying and compensating the Doppler shifts in the receiver \cite{Babar2016, Nguyen2019}. Therefore, \textcolor{black}{the system performance  should be evaluated under the Doppler effects. In this paper, we focus on analyzing the system capacity because it} is an important factor to design components in a communication system.

UWA-OFDM system capacity studies are still an open area up to now as a consequence of lacking the standard channel models \cite{Wang2017, Naderi2017}. Besides, \textcolor{black}{many studies have ignored the ICI effect in the channel capacity formulation} \cite{Nouri2014, Bouvet2010, Milica2008, Aval2015, Radosevic2009}. For the sake of simplicity, \textcolor{black}{the time-invariant UWA channel model in which the ICI effect is thus not occurred  has been used in \cite{Nouri2014, Bouvet2010, Milica2008} to analyze system capacity. The authors in \cite{Aval2015,Radosevic2009} have considered the time-variant channels to analyze the channel capacity versus the SNR, which means that the ICI effect has not been taken into account.} Furthermore, the ICI is also computed by the time correlation function converted from the Doppler spectrum form of the UWA channel, which is assumed to be Jack, uniform or two-path spectrum \cite{Yeli2001, Yeli2005, Nguyen2019}. These assumptions may not be applicable to UWA channels due to the complicated time varying characteristics of the shallow UWA propagation environments. In addition, the UWA channels are considered as non stationary channels such that the conversion between the Doppler spectrum and the time correlation function should be accompanied by certain conditions. \textcolor{black}{In \cite{Bouvet2020}, the capacity of UWA-OFDM has been derived from the SINR which is considered to be the same for all subcarriers. However, the UWA-OFDM system is inherently a wideband system \cite{Morozs2020, Milica2008}, the Doppler shifts over subcarriers are thus different from each other. Consequently, it is in need to investigate the channel capacity of UWA systems under the ICI impacts over every subcarriers.}

\textcolor{black}{Over the past few decades, although a large variety of UWA channel models have been proposed, there is still no typical model that can be applied for all UWA channels because of differences in geographical areas, weather conditions, and seasonal cycles \cite{Aval2015}. The \textcolor{black} {statistical characteristics} of shallow UWA channels are greatly affected by the distribution of the scatterers on the surface and the bottom which results in the different delays and Doppler frequency shifts of the transmit signals. The geometry-based UWA channel model in \cite{Zajic2011} has been derived from computing the number of scatterers and their positions using the wave-guide geometry which does not represent the surface displacement. In \cite{Morozs2020, Widiarti2018, Withamana2017}, the proposed UWA channel models concentrate on analyzing the path loss and the multipath propagation whereas the Doppler effect has not been integrated with the models.} 

In this paper, we use the geometry-based channel model for shallow water with rough surface conditions \cite{Naderi2014} to investigate the quality of the UWA-OFDM system. Particularly,  \textcolor{black}{the scattering points are assumed to be uniformly distributed between the transmitter and the receiver. From the assumption, the pulse response time variation of the channel pattern is obtained. Using this channel impulse response, the SINR of each sub-carrier is derived for the considered system. As aforementioned, the SINR of shallow UWA channels with rough surface transmission system is strongly depended on ICI   effects which is caused by Doppler shifts. We have, in contrast of other works, derived SINR expression of each sub-carrier with considering of the time-variant UWA channel model. We have also used the geometrical scattering model for representing the characteristics of shallow water with rough surface conditions to evaluate the channel capacity.} The channel capacity is further formulated as the superposition of all the single channel capacity from each sub-channel. By analyzing the numerical results, a set of suitable parameters for the considered UWA-OFDM system are found, which includes the number of subcarriers, signal bandwidth and the length of the OFDM symbol. It should be noticed that these are important parameters and they need to be determined appropriately in order to eliminate ISI  while limiting the ICI effects.

The rest of the paper is organized as follows: Section~\ref{Sec:ChannelModel} presents the UWA geometry-based channel model which is used to derive the time-variant channel impulse response and the channel transfer function. The calculations of the SINR expression and channel capacity for the considered UWA-OFDM system using this channel model is presented in detail in Section~\ref{Sec:ChannelCapacity}. Section~\ref{Sec:ResultsandDiscussion} shows the simulation results together with discussions. Finally, Section~\ref{sec:con} draws the main conclusions of this paper. 

\section{\bfseries THE UNDERWATER ACOUSTIC \\GEOMETRY-BASED CHANEL MODEL }
\label{Sec:ChannelModel}
\begin{figure}[t]
   \centering
   \includegraphics[width=0.5\textwidth]{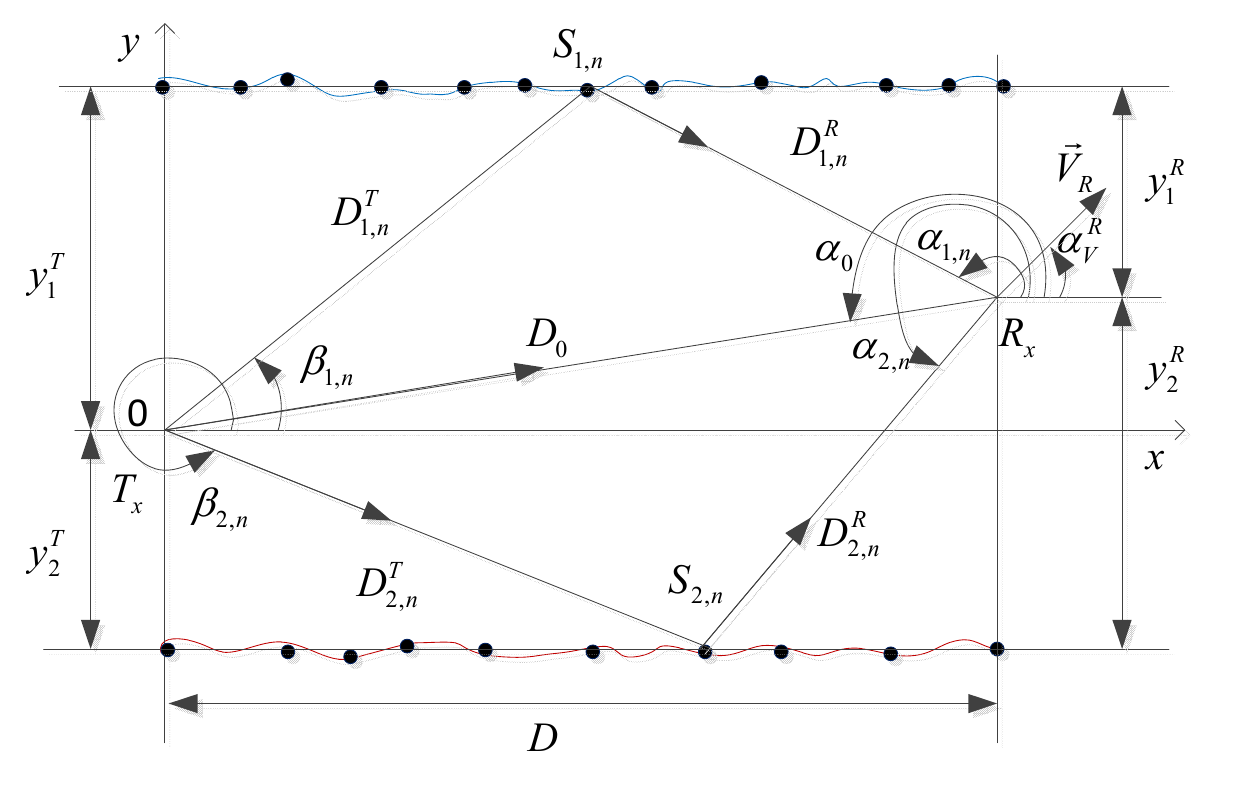}
   \vspace{-0.8cm}
   \caption{The Geometry-Based channel model for a shallow underwater acoustic channel.}
    \label{Fig1_geo}
\end{figure}

The geometry-based channel model \cite{Naderi2014} is illustrated in Fig.~\ref{Fig1_geo}, where $D$ denotes the distance between the transmitter (Tx) and receiver (Rx). The scatterers $S_{i,n} (n = 1, 2, \dots, N_i; i = 1, 2)$  are assumed to be randomly distributed on the surface $(i=1)$ and the bottom $(i=2)$ of a shallow-water environment. The symbols $\alpha_{i,n},\beta_{i,n}$, $(\beta_{i,n} \neq 0;\alpha_{i,n}\neq\pi)$ are the angle-of-departure (AOD) and the angle-of-arrival (AOA) of the $n$-th path, respectively. \textcolor{black}{In the below subsections, the UWA channel parameters are derived with reference to \cite{Naderi2014}.}

\subsection {\bfseries The Channel Impulse Response} 

The time-variant channel impulse response (TVCIR) $h\left( \tau ,t \right)$ of the shallow UWA environment is composed of three components, which is given by \cite{Naderi2014}
\begin{equation} \label{Eq.TVCIR}
    h\left( \tau ,t \right)=\sum\limits_{i=0}^{2}{{{h}_{i}}\left( \tau ,t \right)}.
\end{equation}
In Eq.~\eqref{Eq.TVCIR}, the line-of-sight (LOS) component is denoted by ${{h}_{0}}\left( \tau ,t \right)$, whereas ${{h}_{1}}\left( \tau ,t \right)$ and ${{h}_{2}}\left( \tau ,t \right)$ stand for the scattered components from the surface and the bottom, respectively.
The LOS part ${{h}_{0}}\left( \tau ,t \right)$ is specified by
\begin{equation}
  {{h}_{0}}\left( \tau ,t \right)=\sqrt{\frac{{{c}_{R}}}{1+{{c}_{R}}}}{{A}_{s}}({{D}_{0}}){{A}_{a}}({{D}_{0}})\text{ }{{e}^{j(2\pi {{f}_{0}}t+{{\theta }_{0}})}}\delta (\tau -{{\tau }_{0}}),
\end{equation}
in which $c_R$, $\tau_0$, $f_0$ and $\theta_0$ denote the Rice factor, the propagation delay, the Doppler frequency, and the phase shift of the LOS path, respectively. 
The function $A_s(D_0)$ is the propagation loss coefficient due to spherical spreading, which can be obtained by
\begin{equation}
A_s(D) = \frac{1}{D},
\end{equation}
where $D$ denotes the the total propagation distance in meter. For the LOS path, the distance is defined by 
\begin{equation}
D_0=\sqrt {D^2+(y_1^T-y_1^R)^2}.
\end{equation}
The absorption loss coefficient $A_a(D)$ is given by 
\begin{equation}\label{A_a}
A_a(D) = 10^{-\frac{D_\beta}{2000}}.
\end{equation}
The parameter $\beta$  in Eq.~\eqref{A_a} is computed as 
\begin{equation} 
    \begin{split}
        \beta = 8.68\times10^3 \left(\frac{S_af_Tf_c^2A}{f_T^2+f_c^2} + \frac{Bf_c^2}{f_T}\right)\times \\
        \times(1-6.54\times10^{-4}P) [dB/km],
    \end{split}
\end{equation}
where $A=2.34\times10^{-6}$ and $B= 3.38\times10^{-6}$, $S_a$ is salinity (in ppt), $f_c$ is the carrier frequency (in $\text{kHz}$), $f_T$ is the relaxation frequency (in $\text{kHz}$) and $T$ is temperature (in $\rm ^{\circ}C$). \textcolor{black}{The symbol} $P$ denotes the hydro-static pressure (in $\rm kg/cm^2$), which is determined by $P = 1.01(1+0.1h)$, where $h$ is the water depth (in meter).
The scattered components $h_1(\tau,t)$ and $h_2(\tau,t)$ of the TVCIR $h(\tau,t)$ are computed by
\begin{equation}\label{Eq.TVCIR_i}
    \begin{split}
        {{h}_{i}}\left( \tau ,t \right)=\frac{1}{\sqrt{2{{N}_{i}}(1+{{c}_{R}})}}\sum\limits_{n=1}^{{{N}_{i}}}{{{A}_{s}}({{D}_{i,n}}){{A}_{a}}({{D}_{i,n}})} & \\
  \times \text{ }{{e}^{j(2\pi {{f}_{i,n}}t+{{\theta }_{i,n}})}}\delta (\tau -{{\tau }_{i,n}}), & \\ 
    \end{split}
\end{equation}
in which $f_{i,n}$, $\tau_{i,n}$, and $\theta_{i,n}$ denote the Rice factor, the propagation delay, the Doppler frequency, and the phase shift of the scattered path, respectively. With reference to Fig.~\ref{Fig1_geo}, the total propagation distance  $D_{i,n}$  can be computed as 
\begin{equation}\label{Eq.Di}
D_{i,n} = \frac{y_i^T}{{\rm sin}(\beta_{i,n})} + \frac{y_i^R}{{\rm sin}(\alpha_{i,n})}.
\end{equation}
The parameters of the UWA channel are initiated by using optimum values of $x_{i,n}$ 
\begin{equation}
x_{i,n}^{\rm opt} = \frac{D}{N_i}\left( n-\frac{1}{2}\right).
\end{equation}
Using $x_{i,n}^{\rm opt}$, the other values   AOA $\alpha_{ i,n}$  and AOD $\beta_{ i,n}$ can be computed as follow
	\begin{eqnarray} \label{Alpha}
	\alpha_{i,n} = \begin{cases}
	\frac{\pi}{2}+\arctan\left(\frac{D-x_{i,n}}{y_1^R}\right),
	&\text{if } 0 \leq x_{i,n}\leq D.\\
	\pi + \arctan \left(\frac{y_2^R}{D-x}\right), &\text{if } 0 \leq x_{i,n}\leq D.
	\end{cases}
	\end{eqnarray}
	\begin{eqnarray}\label{eq:beta}
	\beta_{i,n} = \begin{cases}
	\arctan\left(\frac{y_1^T}{x_{i,n}}\right), &\text{if } 0 \leq x_{i,n}\leq D.\\
	\frac{3\pi}{2}+\arctan\left(\frac{x_{i,n}}{y_2^T}\right), &\text{if } 0 \leq x_{i,n} \leq D.
	\end{cases}
	\end{eqnarray}
Substituting  $\alpha_{ i,n}$  and  $\beta_{ i,n}$ in Eq.~\eqref{Eq.Di} allows us to compute $D_{i,n}$. The Doppler frequencies is computed by $f_{i,n} = f_{D,\text{max}}\text{cos}(\alpha_{ i,n}-\alpha_v^R)$, where $f_{D,\text{max}}$ stands for the maximum Doppler frequency. The propagation delays can be obtained by $\tau_{i,n} = D_{i,n} /c_s$. The phase shift $\theta_{i,n}$ is assumed to be uniformly distributed over the range $(-\pi,\pi]$. Finally, the scattered components $h_1(\tau,t)$ and $h_2(\tau,t)$ of the TVCIR $h(\tau,t)$ in Eq.~\eqref{Eq.TVCIR_i} are derived.

\subsection{\bfseries Channel Transfer Function}

For further analysis of the UWA-OFDM system, the time-variant channel transfer function (TVCTF) $H(f,t)$ needs to be derived by taking the Fourier transform of the TVCIR $h(\tau,t)$, which can be expressed by
\begin{equation}
  H\left( f,t \right)=\sum\limits_{i=0}^{2}{{{H}_{i}}\left( f,t \right)},
\end{equation}
where 
\begin{equation}
    \begin{split}
  {{H}_{0}}\left( f,t \right)=\sqrt{\frac{{{c}_{R}}}{1+{{c}_{R}}}}{{A}_{s}}({{D}_{0}}){{A}_{a}}({{D}_{0}}) & \\
  \text{              }\times \text{ }{{e}^{j\left[ 2\pi ({{f}_{0}}t-f{{\tau }_{0}}) \right]}}, & \\ 
\end{split}
\end{equation}
and
\begin{equation}
    \begin{split}
         {{H}_{i}}\left( f,t \right)=\frac{1}{\sqrt{2{{N}_{i}}(1+{{c}_{R}})}}\sum\limits_{n=1}^{{{N}_{i}}}{{{A}_{s}}({{D}_{i,n}}){{A}_{a}}({{D}_{i,n}})} & \\
  \text{ }\times {{e}^{j\left[ 2\pi ({{f}_{i,n}}t-f{{\tau }_{i,n}})+{{\theta }_{i,n}} \right]}} & \\ 
    \end{split}
\end{equation}
\textcolor{black}{for $i=1,2$.}
\section{\bfseries CHANNEL CAPACITY ANALYSIS UNDER INTERFENCE EFFECTS} \label{Sec:ChannelCapacity}

\textcolor{black}{ This section uses the geometry-based UWA channel simulation model to analyze the ICI effect in the UWA-OFDM system. The SINR of each sub-carrier has been formulated in terms of the time-variant channel transfer function (TVCTF) $H(f,t)$ of the UWA channel model. The capacity estimation of UWA-OFDM system has been analyzed by using the results of SINRs.}

\subsection{\bfseries SINR Computation}

The OFDM base-band signal is given as 
\begin{equation} 
x[t_n] = \frac{1}{\sqrt{N}} \sum_{k=0}^{N-1}X[f_k]e^{j 2 \pi n k/N},
\end{equation}
where $N$ is the number of sub-carriers, $X_k$ denotes the $k^{\rm th}$ data-modulated sub-carrier in the OFDM symbol. The OFDM signal at the receiver side is represented as
\begin{equation}\label{eq7}
\widehat{x}[t_n] = \frac{1}{\sqrt{N}} \sum_{k=0}^{N-1}H[f_k,t_n] X[f_k]e^{i2\pi nk/N}+w[t_n],
\end{equation}
where  $H [f_k, t_n]$ stands for the TVCTF of the UWA channel and $w [t_n]$ denotes the ambient noise in the UWA communication system. After FFT at the receiver, the signal $\widehat{X}[f_k]$ in frequency domain can be expressed as
\begin{equation}\label{eq8} 
\widehat{X}[f_k] = \frac{1}{\sqrt{N}} \sum_{n=0}^{N-1} \widehat{x}[t_n] e^{-i2\pi nk/N}.
\end{equation}
By using $\widehat{x}[t_n]$ from \textcolor{black}{Eq.~\eqref{eq7} to Eq.~\eqref{eq8}}, the $\widehat{X}[f_k]$ is given as in \textcolor{black}{Eq.~\eqref{label23}}, where  $S[f_k]$, $I [f_k]$, $W [f_k]$ denote the desired signal, the interference signal, and the frequency domain of ambient noise $w [t_n]$, respectively, which are computed by
\begin{equation}\label{eq10}
S[f_k] = \frac{1}{N} \sum_{n=0}^{N-1}H[f_k,t_n] X[f_k],
\end{equation}
\begin{equation}\label{I_fk}
I[f_k] = \frac{1}{N} \sum_{\substack{m=0\\m\neq k}}^{N-1} \sum_{n=0}^{N-1} H[f_m,t_n] e^{ j 2 \pi (m-k)n/N} X[f_m],
\end{equation}
and 
\begin{equation}
W[f_k] = \frac{1}{\sqrt{N}} \sum_{n=0}^{N-1} w[t_n] e^{-j 2 \pi n k /N }.
\end{equation}
\textcolor{black}{Assuming the Signal to Noise Ratio of the $k^{\rm th}$ subcarrier in the absence of ICI is denoted by the ${\rm SNR}[f_k]$, which is given by}
\begin{equation}\label{SNR}
\text{SNR}[f_k] = \frac{P_\text{S}[f_k]}{P_\text{N}[f_k]},
\end{equation}
\textcolor{black} {where $P_\text{S}[f_k]$ is the receiver power and $P_\text{N}[f_k]$ stands for the noise power at the $k^{\rm th}$ subcarrier. Using \eqref{eq10} and \eqref{I_fk}, the desired signal power $P_\text{D}[f_k]$ and the ICI power $P_\text{I}[f_k]$  of the $k^{\rm th}$ subcarrier can be obtained by}
\begin{equation}\label{P_des}
P_\text{D}[f_k] = \frac{P_\text{S}[f_k]}{N^2} \left(\sum_{n=0}^{N-1}H[f_k,t_n]\right)^2,
\end{equation}
and
\begin{equation}\label{P_I}
P_\text{I}[f_k] = \frac{P_\text{S}[f_k]}{N^2}\left(\sum_{\substack{m=0\\m\neq k}}^{N-1} \sum_{n=0}^{N-1} H[f_m,t_n] e^{ j 2 \pi (m-k)n/N}\right)^2.
\end{equation}
\textcolor{black}{The SINR of the kth sub-carrier is given by}
\begin{equation}
\text{SINR}[f_k] = \frac{P_\text{D}[f_k]}{P_\text{I}[f_k]+P_\text{N}[f_k]}.
\end{equation}
Using \textcolor{black}{Eq.~\eqref{SNR}, Eq.~\eqref{P_des}, and Eq.~\eqref{P_I},} the ${\rm SINR}$ of the $k^{\rm th}$ subcarriers can be obtained as in Eq.~\eqref{eq9}.
\begin{figure*} [t]
		 \vbox{\begin{align} \label{label23}
		\widehat{X}[f_k] &= \frac{1}{N} \sum_{m=0}^{N-1}\left[ \left( \sum_{n=0}^{N-1} H[f_m,t_n] e^{j 2 \pi (m-k)} \right) X[f_m]\right] + W[f_k]
= S[f_k] + I [f_k] + W [f_k].
		\end{align}
		}
\end{figure*}
\begin{figure*} [t]
	\vbox{\begin{align} \label{eq9}
		\mathrm{SINR}[f_k] &= \frac{{\vert \frac{1}{N} \sum_{n=0}^{N-1} H[f_k,t_n] \vert }^2 }{\Biggl| \frac{1}{N} \sum_{m=0, m\neq k}^{N-1} \sum\limits_{n=0}^{N-1} H[f_m,t_n] e^{ j 2 \pi (m-k)n/N} \Biggr|^2  + \frac{1}{\text{SNR}[f_k]}}.
		\end{align}
	}
\hrule
\end{figure*}

\subsection{\bfseries Capacity Analysis} 

The channel capacity of $k^{\rm th}$ sub-carrier of an UW-OFDM system is computed from the ${\rm SINR}[f_k]$ as follows
\begin{equation}
C_k = \Delta f {\rm log}_2(1 + {\rm SINR}[f_k]).
\end{equation}
Suppose that each sub-carrier carries data, the total channel capacity of UWA-OFDM system is calculated by
\begin{equation}\label{C_SINR}
C_{\rm SINR} = \frac{T_S}{T_S + T_G} \Delta f \sum_{k=0}^{N-1} {\rm log}_2(1 + {\rm SINR}[f_k]),
\end{equation}
where $T_S, T_G $ are the symbol duration and the guard length \textcolor{black}{of the UWA-OFDM system}, respectively. 

In UWA \textcolor{black}{communication systems},  the spectral efficiency $C/B$ is a vital parameter in system performance evaluation because of the limited bandwidth. \textcolor {black} {Substituting }$\Delta f = B/N$ into Eq.~\eqref{C_SINR}, we obtain the spectral efficiency $C/B$ as below
\begin{equation}\label{C_B}
\frac{C}{B} = \frac{T_S}{T_S + T_G}\times \frac{1}{N} \times \sum_{k=0}^{N-1} {\rm log}_2(1 + {\rm SINR}[f_k]).
\end{equation}

\textcolor{black}{Observing Eq.~\eqref{C_B}, several constraints need to be considered in determining} OFDM transmission parameters to optimize the spectral efficiency. \textcolor{black}{Assuming that the guard length $T_G$ is chosen to be equal to the maximal delay spread $\tau_{\rm max}$ to remove the ISI noise}. With \textcolor{black}{a given} number of sub-carriers $N_c$, the bandwidth efficiency coefficient $\beta =T_S/(T_S+T_G)$  \textcolor{black}{is increased with a larger symbol duration $T_S$}. However, \textcolor{black}{the larger $T_S = N/B$ makes the sub-carrier spacing $\Delta f=B/N$ decrease. Consequently,} the ICI \textcolor{black}{effect} will be stronger \textcolor{black}{that results in degrading} $\rm SINR$. \textcolor{black}{Therefore, the set of parameters $ {T_S, N_c, B}$ should be considered thoughtfully to meet the quality requirements} of the system.

The \textcolor{black}{numerical results of }$\mathrm{SINRs}$ and system capacity \textcolor{black}{in} section IV will evaluate the appropriate system parameters, \textcolor{black}{including of} $T_S$, $B$, and $N_c$ to achieve optimal spectral efficiency for the UWA-OFDM system with ICI effects.

\section{\bfseries RESULTS AND DISCUSSIONS} \label{Sec:ResultsandDiscussion}

\subsection{\bfseries Simulation Setting}
The results in surveying system performance including $\mathrm{SINR}$ and Capacity are averaged over 10 simulations, each run with a time of 10000 OFDM symbols. The main parameters running in the simulation are taken with central carrier frequency $f_c= 30\,\text{kHz}$, ${\rm SNR}=20\,{\rm dB}$ at receiver, the number of subcarriers $N_c$ = 512, 1024 and 2048, and the signal bandwidth varies between  $1\,\text{kHz}$ and $30\,\text{kHz}$.

\subsection{\bfseries SINR Results}
\begin{figure}[t]
    \centering
    \includegraphics[width=0.5\textwidth]{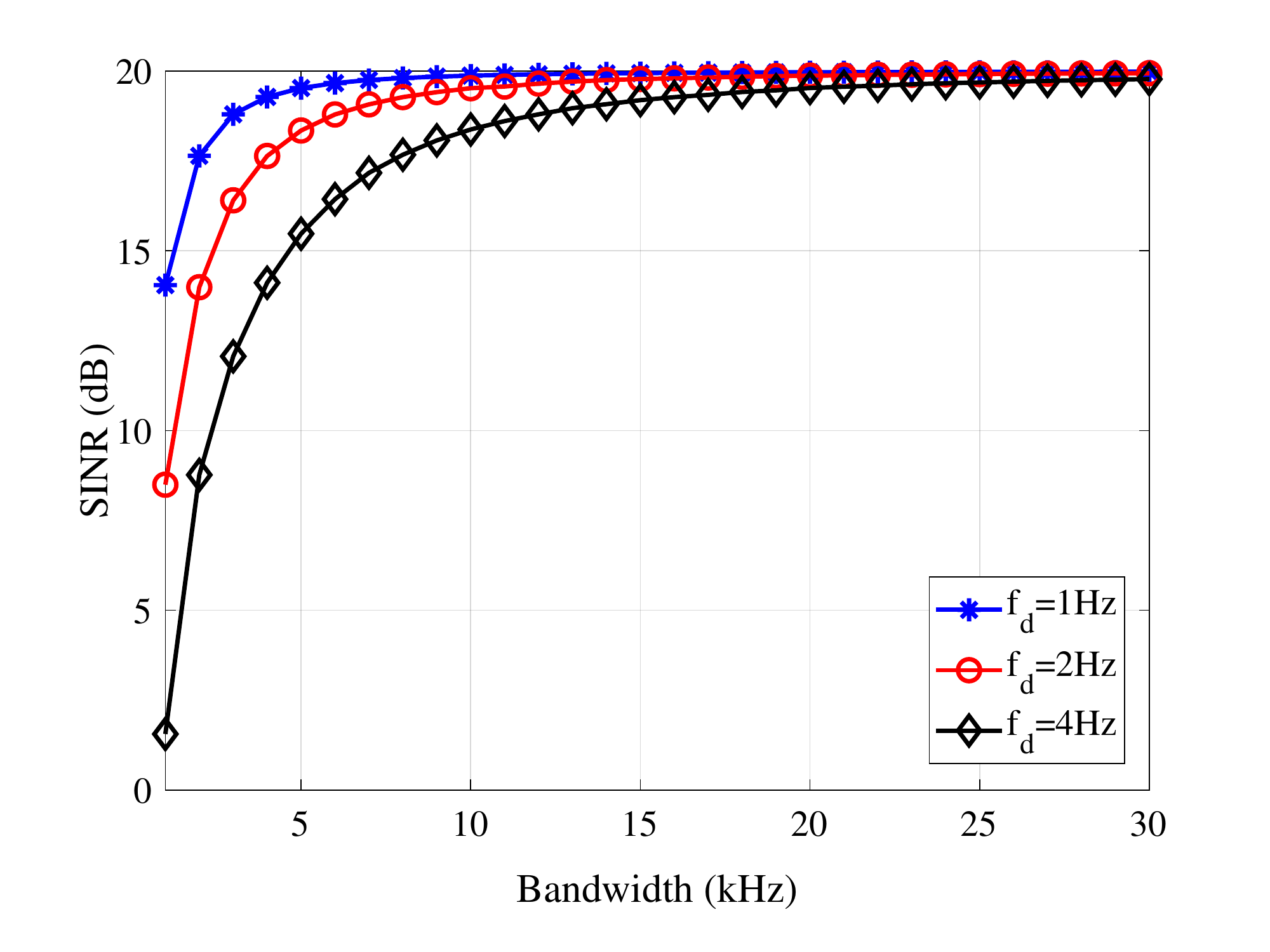}
    \vspace{-0.8cm}
    \caption{SINRs versus signal bandwidth for different Doppler frequencies.}
    \label{Fig1}
\end{figure}
\textcolor{black}{Figure~\ref{Fig1}} shows the ${\rm SINR}$ results of the UWA-OFDM system with the number of subcarriers $N_c =1024$. It is noted that the ${\rm SINR}$ is calculated for each subcarrier and the results in Fig.~\ref{Fig1} is the average of all subcarriers. We see the strong Doppler effect on the ${\rm SINR}$ when the signal bandwidth is small. With \textcolor{black}{a given} number of subcarriers\textcolor{black}{, the smaller signal bandwidth is, the subcarrier spacing $\Delta f=N/B$ is narrower that causes more serious ICI effect and decreasing the ${\rm SINR}$. On the contrary, when increasing the signal bandwidth, the larger carrier spacing mitigates the ICI effect. If  the bandwidth $B$ is large enough, the ICI effect can be neglected and the SINR results approach to the SNR=20\,dB.}
	
From the SINR results in Fig.~\ref{Fig1}, it is observed that with a bandwidth greater than 10\,$\text{kHz}$, the \textcolor{black}{ICI effect for the Doppler frequencies of} $f_d$ = 1\,$\text{Hz}$ and 2\,$\text{Hz}$ is negligible. \textcolor{black}{For the case of} $f_d$ = 4\,$\text{Hz}$, a bandwidth greater than 20\,$\text{kHz}$ is required to significantly reduce the \textcolor{black}{ICI} effect. \textcolor{black}{Therefore,} depending on the hardware capabilities of the system, \textcolor{black}{the impact of ICI on system performance can be significantly reduced if a wide bandwidth is chosen.}
\begin{figure}[t]
    \centering
    \includegraphics[width=0.5\textwidth]{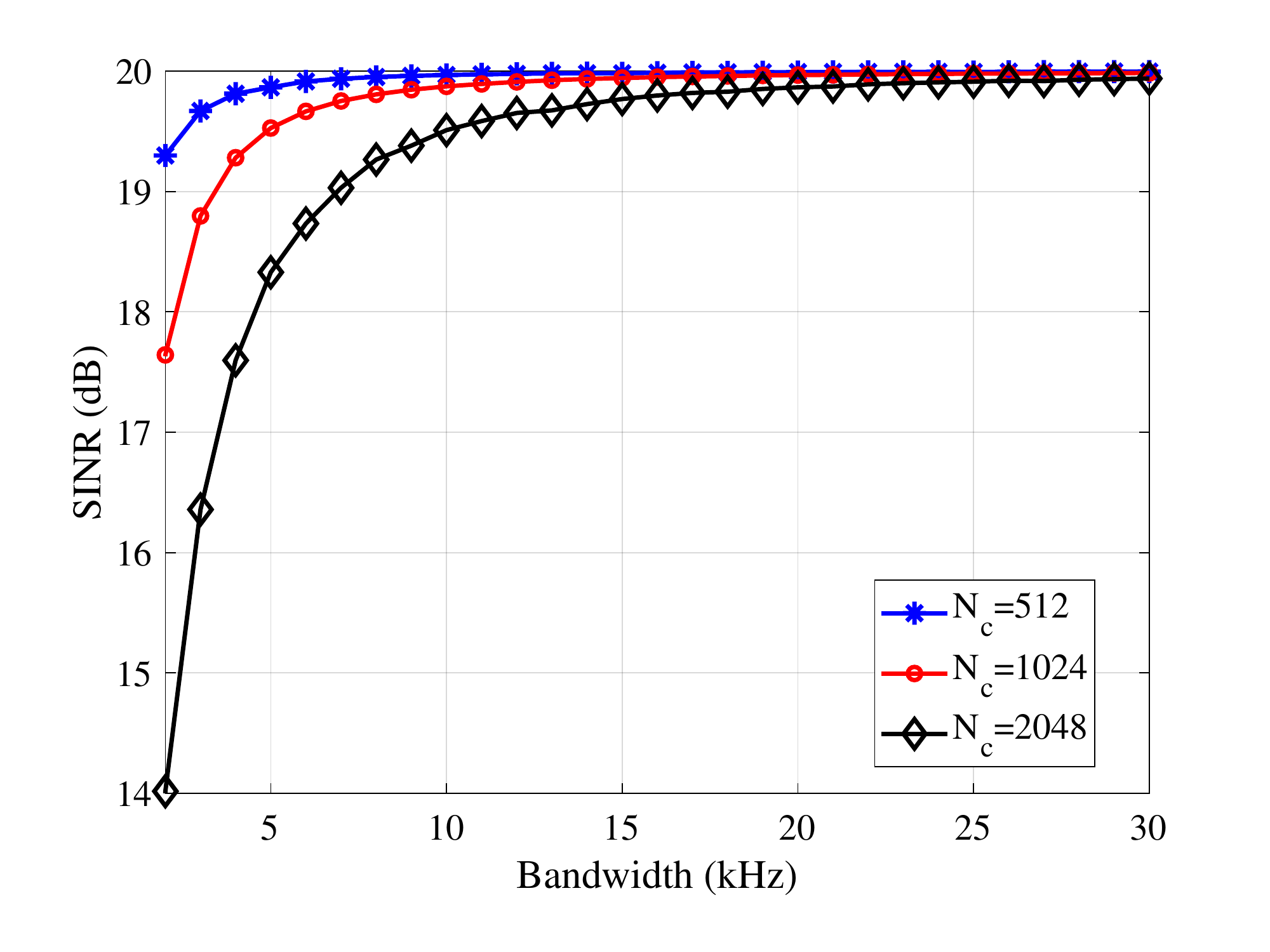}
    \vspace{-0.8cm}
    \caption{SINR versus signal bandwidth for different numbers of subcarriers ($f_d$= 1$\text{Hz}$).}
    \label{Fig2}
\end{figure}

Another aspect \textcolor{black} {which should be considered in UWA-OFDM system design} is the number of subcarriers $N_c$. \textcolor{black}{For a given} bandwidth, the subcarrier spacing is narrower with the larger $N_c$. \textcolor{black}{Consequently, the ICI effect on the OFDM system is more severe and then the SINR decreases. As the results shown in Fig. \ref{Fig2}, the SINR is lower for the case of larger number of subcarriers $N_c$. Using these results, the appropriate values of bandwidth $B$ and the number of sub-carrier $N_c$ can be determined to achieve a
required SINR of the UWA-OFDM system.} \textcolor{black} {For even very small value of Doppler shift $f_d=1\,\text{Hz}$,} the maximum number of subcarriers of $N_c=1024$ and the minimum bandwidth of $10\,\text{kHz}$ \textcolor{black}{should be selected to avoid the ICI effect}. For $N_c=2048$, the minimum bandwidth is \textcolor{black}{required to be} greater than $20\,\text{kHz}$. However, \textcolor{black}{on one hand, a small number of subcarriers results in mitigating the Doppler effect; on the other hand, it makes the efficiency of the spectrum and also the system capacity decrease.} The next section will evaluate the capacity to determine the appropriate number of subcarriers for the UWA-OFDM system.
\subsection{\bfseries Capacity Results}
\begin{figure}[t]
    \centering
    \includegraphics[width=0.5\textwidth]{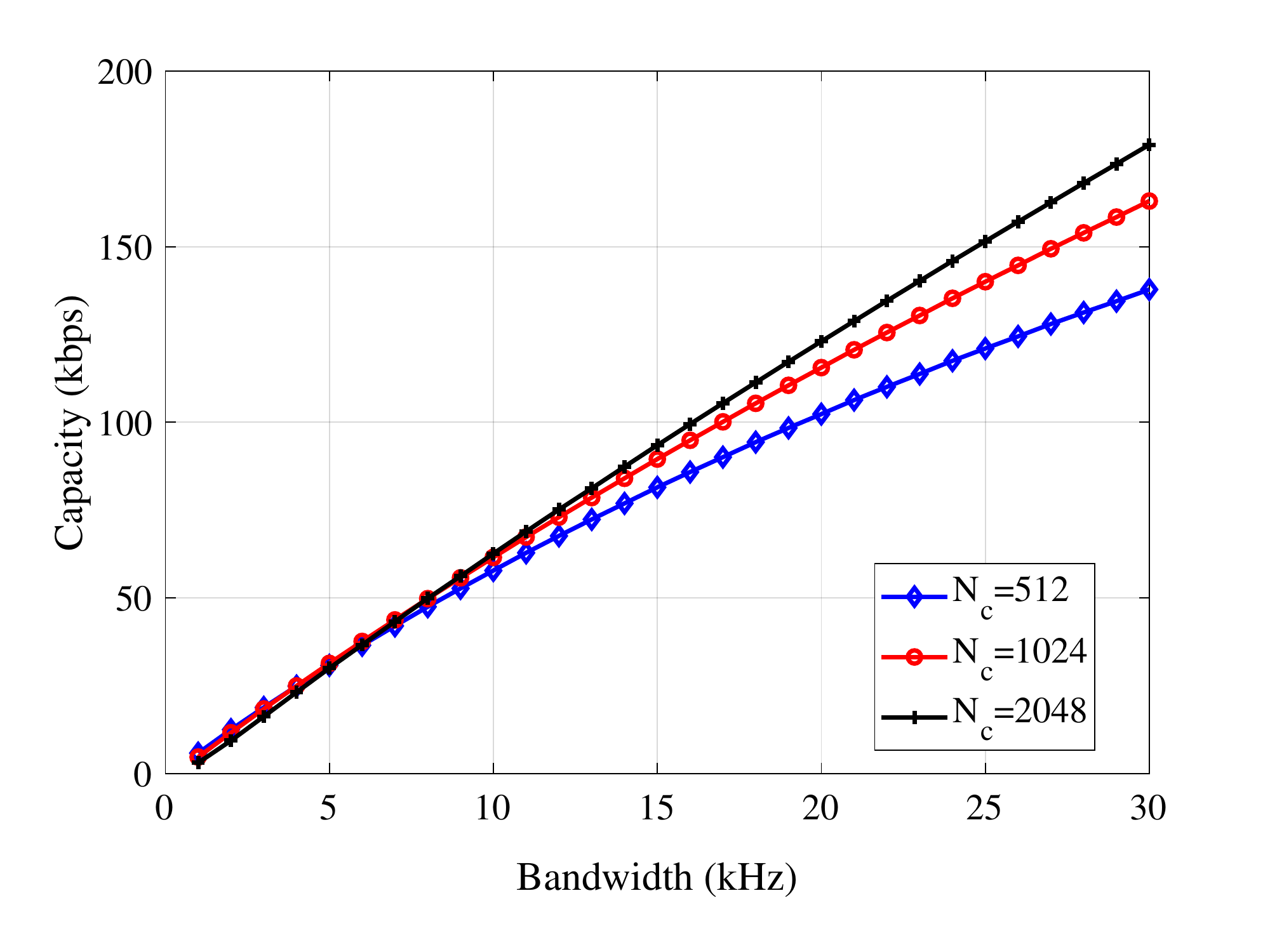}
    \vspace{-0.8cm}
    \caption{System capacity $C\,{\rm (Kbps)}$ versus bandwidth for different numbers of subcarriers ($f_d = 1\text{Hz}$).}
    \label{Fig3}
\end{figure}
The system capacity of the UWA-OFDM system \textcolor{black}{versus} the signal bandwidth for different numbers of subcarriers (with $f_d = 1\text{Hz}$) is shown in Fig.~\ref{Fig3}. \textcolor{black}{For the bandwidth range of $B<10\,\text{kHz}$}, the system capacity is almost same value for the different numbers of subcarriers of 512, 1024, and 2048. In this case, the lowest number of subcarriers $N_c=512$ should be chosen to \textcolor{black}{reduce complexity of receiver.} \textcolor{black}{The reason can be explained by using Eq.~\eqref{C_SINR}. For the narrow bandwidth (i.e. less than 10 $\text{kHz}$ for the considered case), the increase in number of sub-carriers leads to the decrease in the SINR as shown in Fig.~\ref{Fig2}. However, that makes the OFDM symbol $T_S= N/B$ be larger. As a 	result, the bandwidth efficiency $\beta = T_S/ (T_S + T_G)$, in which $T_G = \tau_{max}$ is fixed, will be increased. Therefore, if $N_c$ increases, the bandwidth efficiency will increases, while the SINR decreases. From this argument along with Eq.~\eqref{C_SINR}, we observed also the capacity remains unchanged for the larger numbers of subcarriers as shown in Fig.~\ref{Fig3} with the bandwidth range $B<10\,\text{kHz}$. In other words, with the limited bandwidth of UWA channels, it may be impossible to increase the capacity by simply increasing the number of sub-carriers.}

In a similar way, for the larger bandwidth of range from $10\,\text{kHz}$ to $15\,\text{kHz}$, one should choose $N_c=1024$ to ensure that the system capacity is not significantly reduced, but to limit the ICI effect. For a bandwidth of more than $20\,\text{kHz}$, $N_c = 2048$ is suitable.
\begin{figure}[t]
    \centering
    \includegraphics[width=0.5\textwidth]{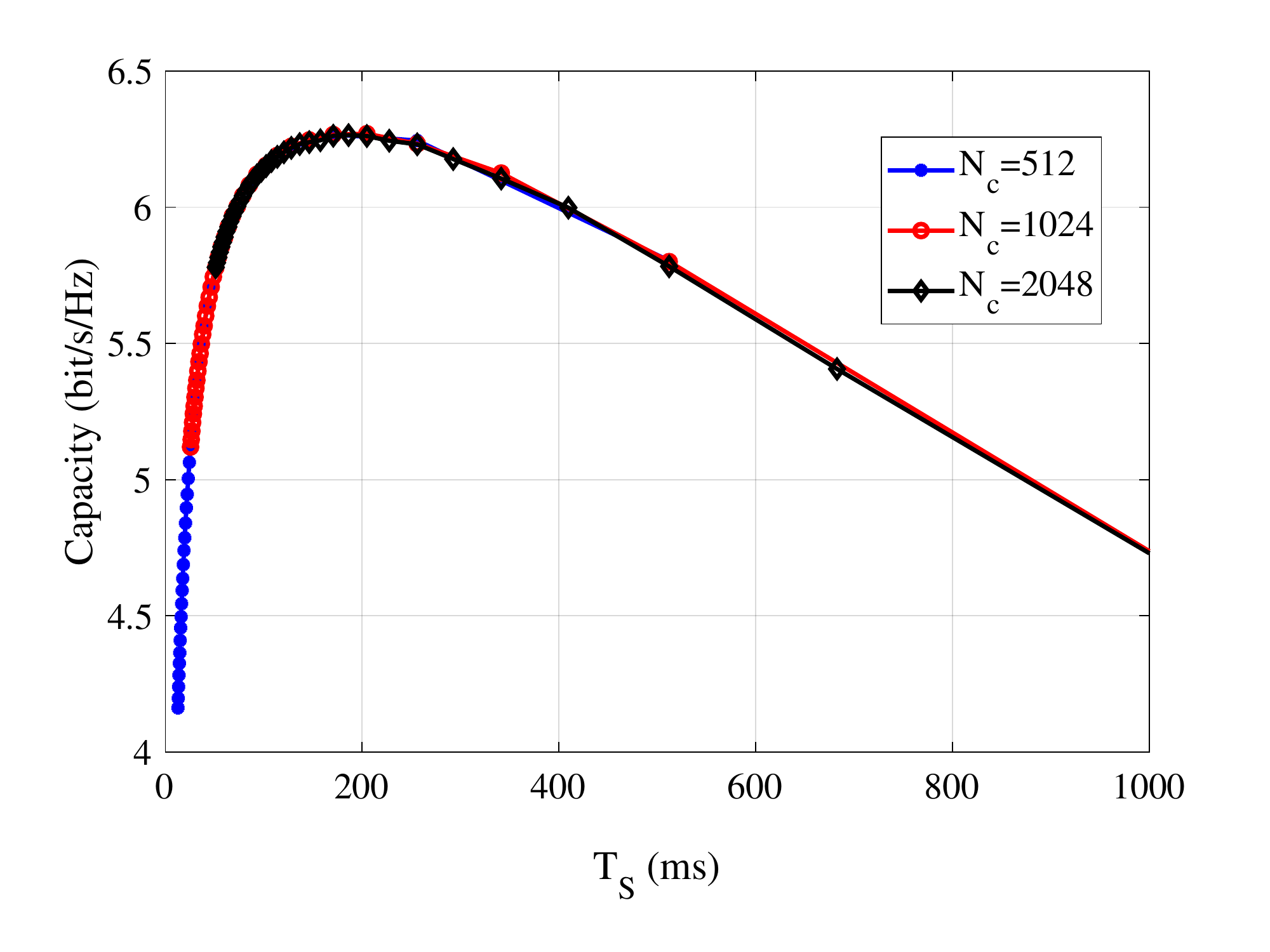}
    \vspace{-0.8cm}
    \caption{Spectral efficiency $C/B\,{\rm (b/s/Hz)}$ versus symbol length $T_S$ for different numbers of subcarriers.}
    \label{Fig4}
\end{figure}

Besides the system capacity $C\,{\rm (Kbps)}$, the spectrum efficiency $C/B\,{\rm (b/s/Hz)}$ is also an important factor in the OFDM system design due to the limited bandwidth of UWA channels. Figure~\ref{Fig4} \textcolor{black}{shows the results of} the spectrum efficiency $C/B\,{\rm (b/s/Hz)}$ versus the symbol duration $T_S$ for different numbers of subcarriers for the case of $f_d=1\,\text{Hz}$. \textcolor{black}{For the small symbol duration range of $T_S < 204.8{\rm ms}$, the larger $T_S$ results in the higher spectral efficiency $C/B$. This is because the larger $T_S$ makes the higher bandwidth efficiency coefficient $\beta$. Consequently, the spectral efficiency $C/B$ will be increased as shown in Eq.~\eqref{C_B}. On the contrary, the increase of $T_S$ will get the decrease of the spectral efficiency $C/B$ for the larger symbol range $T_S > 204.8\,{\rm ms}$. The reason is that the larger $T_S$ (i.e. the smaller subcarrier spacing $\Delta_f$) makes the ICI effect more serious on the UWA channel. Hence, $T_S > 204.8$~ms results also in the decrease of both SINR and the spectral efficiency $C/B$ as shown in Eq.~\eqref{C_B}.} As observed in Fig.~\ref{Fig4}, the spectral efficiency  all achieves the maximal value $C/B_{\rm max} = 6.265\,{\rm(b/s/Hz)}$  at $T_S = 204.8\,{\rm ms}$. Using this result, we can determine the optimal bandwidth for each different number of subcarriers $N_c$.

\begin{figure}[t]
    \centering
    \includegraphics[width=0.5\textwidth]{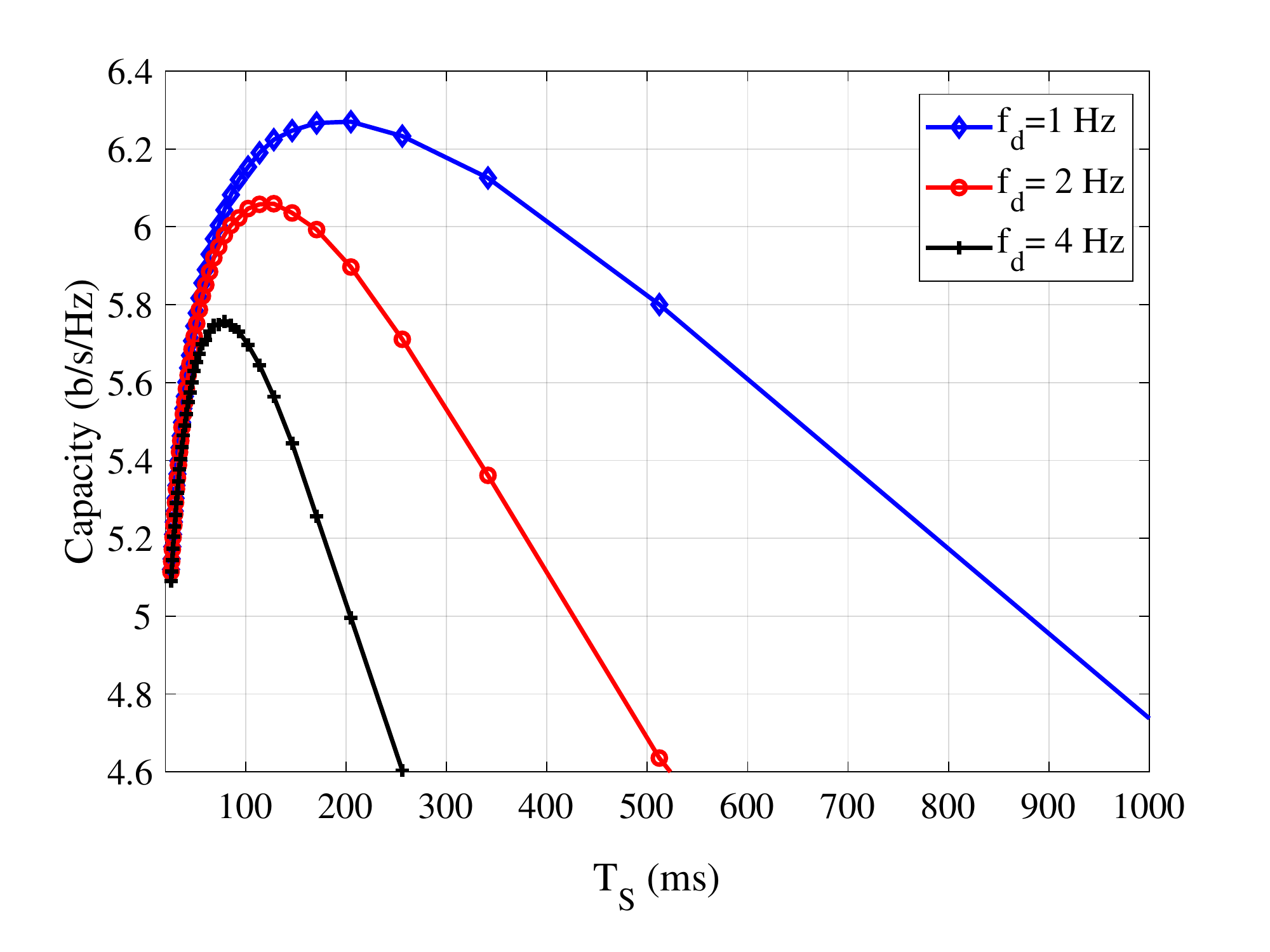}
    \vspace{-0.8cm}
    \caption{Spectral efficiency $C/B\,{\rm (b/s/Hz)}$ versus symbol length $T_S$ for different Doppler frequencies.}
    \label{Fig5}
\end{figure}
Moreover, depending on the Doppler shift $f_d$, the appropriate $T_S$ should be chosen to achieve the maximum $C/B$. Figure \ref{Fig5} \textcolor{black}{depicts} the $C/B$ results versus $T_S$ for different Doppler shifts for the case of $N_c=1024$. When the Doppler frequency increases, the optimal value of $T_S$ to achieve maximum $C/B$ is decreased. The optimal $T_S$ and $B$ values for different Doppler frequencies are shown in Table I. Based on these results, one can determined the parameters ${T_S, B, N_c,}$ for UWA-OFDM systems to not only achieve maximum spectral efficiency and channel capacity but also limit the ICI effect and minimize complexity at the receiver under different transmission conditions.
\begin{table}[]\label{Tab1}
\caption{The UWA-OFDM system parameters for different Doppler frequencies. }
\begin{tabular}{|c|c|c|c|c|c|}
\hline
\multirow{2}{*}{$f_d$ } &
  \multirow{2}{*}{\begin{tabular}[c]{@{}c@{}}$(C/B)_\text{max}$\\    (b/s/Hz)\end{tabular}} &
  \multirow{2}{*}{$T_S$(ms)} &
  \multicolumn{3}{c|}{$B$ (kHz)} \\ \cline{4-6} 
     &       &       & \multicolumn{1}{l|}{N=512} & \multicolumn{1}{l|}{N= 1024} & \multicolumn{1}{l|}{N =2048} \\ \hline
1 Hz & 6.265 & 204.8 & 2.5                         & 5.0                           & 10.0                         \\ \hline
2 Hz & 6.059 & 128.0 & 4.1                         & 8.2                           & 16.4                         \\ \hline
4 Hz & 5.757 & 78.8  & 6.5                         & 13.0                          & 26.0                         \\ \hline
\end{tabular}
\end{table}
\section{\bfseries CONCLUSIONS}\label{sec:con}

This paper has investigated the capacity of the UWA-OFDM system under the impact of ICI effect. Using the  time variant \textcolor{black}{geometry-based} channel model for the shallow water environment, the ${\rm SINR}$ of each sub-carrier has been derived. The channel capacity  is computed from the ${\rm SINR}$ which takes the Doppler effect into account. The numerical results of  ${\rm SINR}$, channel capacity, and spectral efficiency  give us guidance in the UWA-OFDM system design, specifically choosing the transmission parameters including the number of subcarriers, the symbol length, and the signal bandwidth for the different Doppler shifts. 
	
\section*{Acknowledgement}
This study was funded by the Vietnam National Foundation for Science and Technology Development (NAFOSTED) under the project number 102.04-2018.12.

\bibliographystyle{ieeetr}
\bibliography{MyRef2021}
\begin{IEEEbiography}[{\includegraphics[width=1in,height=1.25in,clip,keepaspectratio]{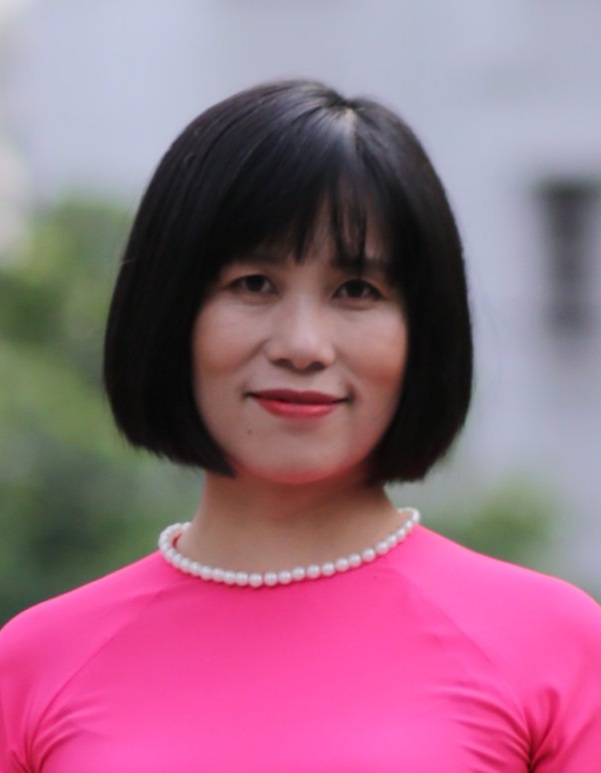}}]{Do Viet Ha}	received the B.S, M.Sc., and Ph.D. degrees in Electronics and Telecommunications Engineering from Hanoi University of Science and Technology (HUST), Hanoi, Vietnam, in 2001, 2007, and 2017, respectively. She is currently working with the Department of Electronics Engineering, University of Transport and Communications, Hanoi, Viet Nam, as a lecturer. Her main areas of research interest are mobile channel modeling, especially underwater acoustic channels, underwater acoustic OFDM systems, and mobile-to-mobile communications
\end{IEEEbiography}
\vspace{-5cm}
\begin{IEEEbiography}[{\includegraphics[width=1in,height=1.25in,clip,keepaspectratio]{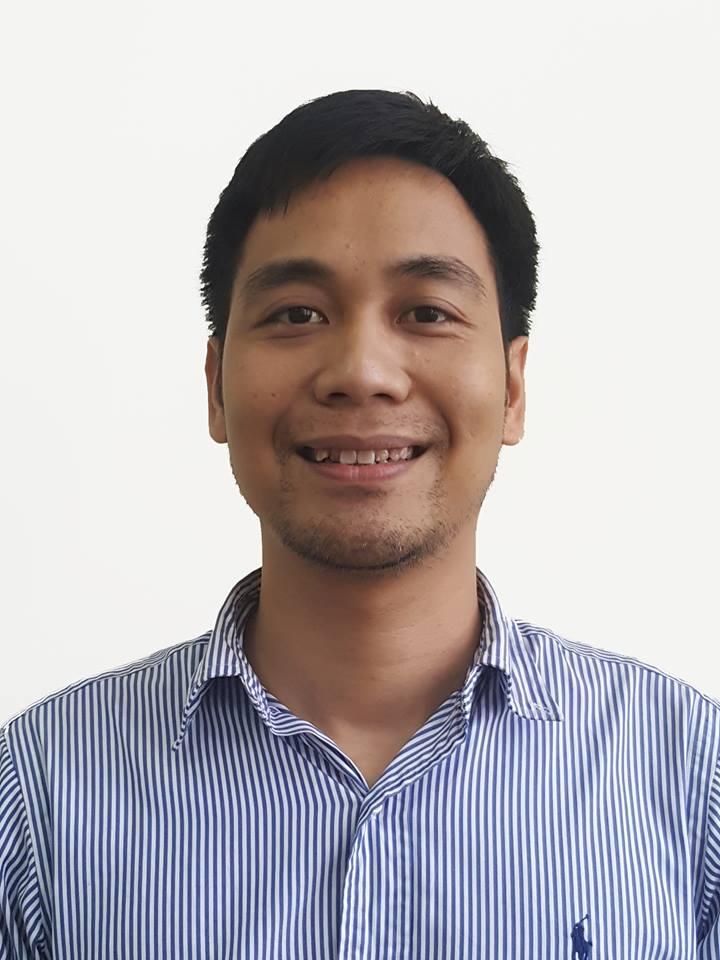}}]{Nguyen Tien Hoa} graduated with a Dipl.-Ing. in Electronics and Communication Engineering from Hanover University. He has worked in the R\&D department of image processing and in the development of SDR-based drivers in Bosch, Germany. He devoted three years of experimentation with MIMOon's R\&D team to develop embedded signal processing and radio modules for LTE-A/4G. He worked as a senior expert at Viettel IC Design Center (VIC) and VinSmart for development of advanced solutions for aggregating and splitting/steering traffic at the PDCP layer to provide robust and QoS/QoE guaranteeing integration between heterogeneous link types, as well as Hybrid Beamforming for Millimetre-Wave 5G systems. Currently, he is a lecturer at the School of Electronics and Telecommunications, Hanoi University of Science and Technology. His research interests are resource allocation in B5G, and vehicular communication systems. 	 
\end{IEEEbiography}
\vspace{-5cm}
\begin{IEEEbiography}[{\includegraphics[width=1in,height=1.25in,clip,keepaspectratio]{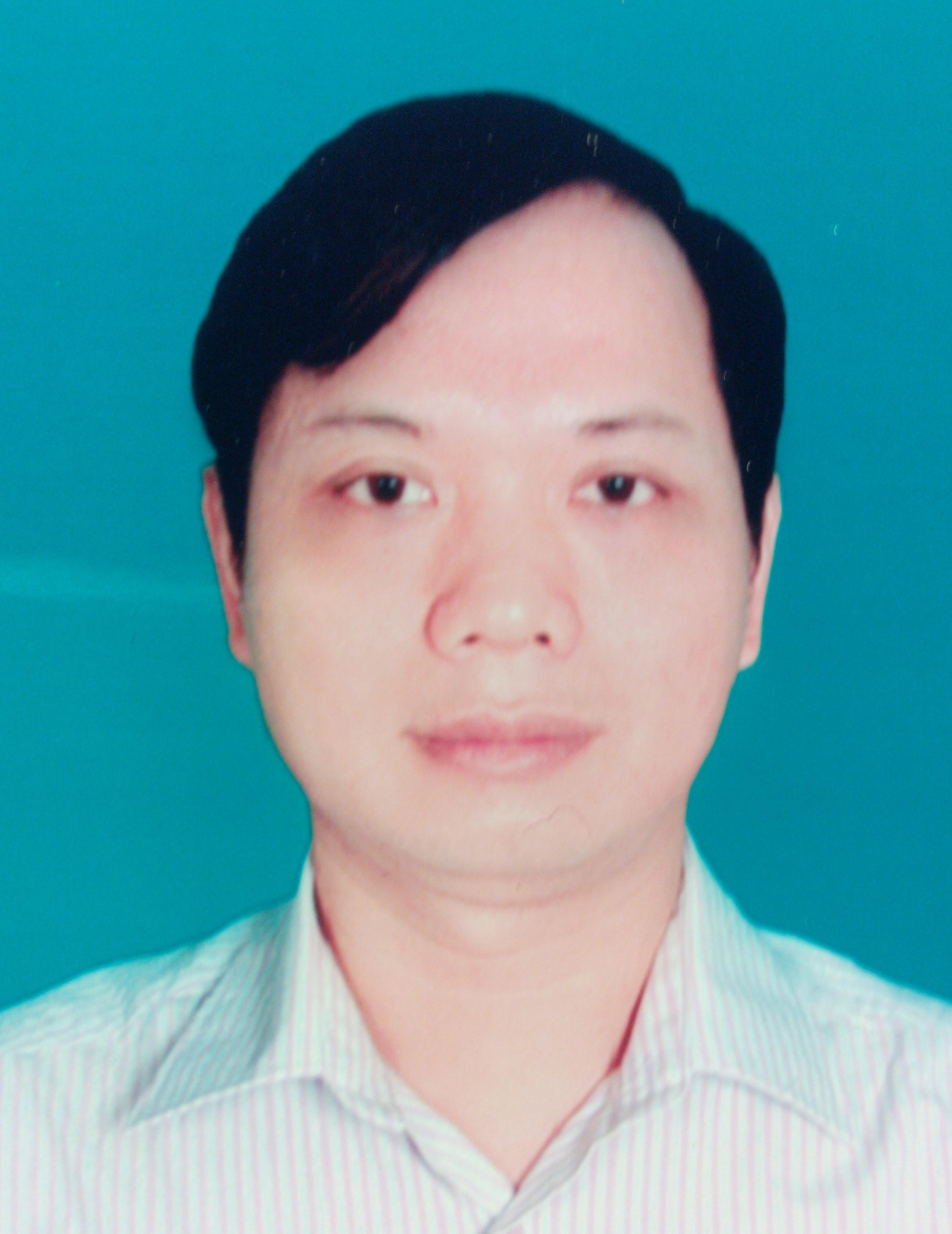}}]{Nguyen Van Duc} received the B.S and M.Sc degrees in Electronics and Telecommunications engineering from Hanoi University of Science and Technology (HUST), Hanoi, Vietnam, in 1995 and 1997, respectively. In 2003, he received the Ph.D. degree in Electronics and Telecommunications Engineering from Leibniz University Hannover, Germany. Since 2006, he has been with the Department of Communications Engineering, HUST, where he is currently an Associate Professor. His current research interests include underwater communications, intelligent transport systems, and cognitive radio networks. 
\end{IEEEbiography}
\end{document}